\documentstyle[preprint,aps]{revtex}

\begin{document}

\preprint{LSUHE No. 141-1993 }
\def\overlay#1#2{\setbox0=\hbox{#1}\setbox1=\hbox to \wd0{\hss #2\hss}#1%
\hskip -2\wd0\copy1}

\title{$SU(2)$ Flux Distributions On Finite Lattices}

\author{Yingcai Peng and Richard W. Haymaker}
\address{Department of Physics and Astronomy, Louisiana State University, Baton Rouge, La 70803-4001, USA}
\maketitle
\begin{abstract}
     We studied $SU(2)$ flux distributions on four dimensional
 Euclidean lattices
with one dimension very large. By choosing the time direction
appropriately we can study physics in two cases:
one is finite volume in the zero temperature limit,
another is finite temperature in the intermediate to
large volume limit. We found that for cases of $\beta > \beta_c$
there is no intrinsic string formation. Our lattices with
$\beta > \beta_c$ belong to the intermediate volume region,
and the string tension in this region
is due to finite volume effects. In large volumes we found
evidence for intrinsic string formation.
\end{abstract}
\pacs{ }

\narrowtext
\section{Introduction}
\label{sec: intro}

There are two complementary ways to obtain approximate
solutions of strong coupling QCD. One is to take 3-space to be
a finite volume torus and obtain semi-analytic
solutions\cite{l,lm,vk} of an effective Hamiltonian.
The second is via lattice simulations.
Confinement can be studied by each method in
limited domains.   In small physical volumes the fields are very
rigid and the problem can be treated using a variational method
applied to a small number of dynamical variables.
In this domain the string tension is found to be zero.
This result is consistent with asymptotic freedom
since only short wavelength modes occur.
At intermediate volumes there is a
clear signal for string tension and further that
it is a consequence of
a tunneling amplitude between the vacuua that are degenerate for
small volumes.
In this domain, lattice methods are also accessible and are in
good agreement for quantities such as glueball masses and
string tension. Lattice calculations can take over to study
larger volumes where semi-analytic methods become prohibitive.

The existence of string tension in finite volumes does not
imply confinement. Clearly if the the volume is not large enough
to allow the fields to spread out, the finite box itself may be
responsible for the linearly rising potential energy between quarks.
Global studies have left open the question
of the volume at which intrinsic confinement takes over\cite{bb}.
In this paper we look at a local quantity,  the flux tube
profile between static quarks as a function
of physical volume in SU(2) lattice gauge theory in order to
elucidate this question.

The physical size of the box is characterized by a dimensionless
variable
\begin{equation}
 z_g\equiv m(0^+)L,
\label{e1.1}
\end {equation}
where $m(0^+)$ is the lowest glueball mass which is the
energy gap or in terms of length it is the inverse of the correlation
length. The length L is the linear size of the box.
M. L\"{u}scher studied QCD in a small box, $z_g \leq 1$, with
periodic boundary conditions. He derived a low-energy effective
Hamiltonian for SU(N) gauge theory in small volumes\cite{l},
i.e. $z_g \leq 1$. Subsequently the lowest energy levels of
SU(2)\cite{lm} and SU(3)\cite{wz}  gauge theories in small volumes
were computed by using this Hamiltonian.  Van Baal and Koller
then found that the crucial tunnelling between degenerate vacuua can
be obtained by imposing appropriate nonperturbative boundary
conditions on the Raleigh-Ritz trial wave functions\cite{vk}.
They extended the calculation of the  SU(2) glueball masses up
to $z_g \approx 5.0$.

Berg and Billoire \cite{bb} carried out a thorough study of
glueball masses,
electric flux states and string tension for intermediate volumes
$(1 \leq z_g \leq 5)$.  They provided a detailed comparison between
their numerical results and the analytic results of Van Baal and Koller.
They chose lattice sizes $N_a \times N_b^2 \times N_c$, with
\begin{equation}
N_a \leq  N_b \ll  N_c.
\label{e1.2}
\end{equation}
By identifying $N_c$ to be the time extent, one can simulate
the zero temperature finite volume, $N_a \times N_b^2 a^3$, field theory,
with the temperature defined as \cite{cb}
\begin{equation}
 T_B=\frac {1}{N_c a} ,
\label{e1.3}
\end{equation}
where $T_B$ is called {\it box temperature} in ref.\cite{bb}, and $a$
is the lattice spacing.

Physically, the choice of the time
direction is related to interpreting Polyakov loop correlations
as the $q\bar q$ potential; then the time direction is
the one in which the Polyakov loop closes. In our study we
follow Berg and Billoire to choose Polyakov loops closed in the
$N_a$ direction and their correlations measured along the
$N_c$ direction. Then the {\it physical temperature} is defined to be
\begin{equation}
 T_p=\frac {1}{N_a a}.
\label{e1.4}
\end{equation}

  In the coming sections we shall use the physical
temperature ($T_p$) interpretation, so we shall drop the subscript $p$,
i.e. $T=T_p$. In the last section we shall relate this
to the box temperature ($T_B$) interpretation.

   After a complete study in the intermediate volume region ($1<z_g<5$),
Berg and Billoire concluded that they did not find
evidence for string formation in this region,
but it is expected to occur in larger volumes.
The question this paper studies is how to understand the string tension
measured in intermediate volumes. Is the string tension just
due to finite size effects? Since
the string tension was calculated from correlations
of Polyakov loops closed in the $N_a$ direction, this can
be considered as measuring the potential energy of a
$q\bar q$ pair in a long rectangular box, with the volume
$V=N_b^2\cdot N_c \cdot a^3$ and at the finite temperature
$T=1/N_a a$, as shown in Fig.~\ref{f1.1}. As we know,
at high temperatures, $T>T_c$, where $T_c$ is the
deconfining temperature, the $q\bar q$ is unconfined
in the infinite volume limit ($N_b$, $N_c \rightarrow \infty$ ).
However, as the transverse size $N_b$ is made small,
i.e. $r/N_b a \ge 1$, one expects the side walls of the box
would squeeze the flux lines of the $q\bar q$ pair to form a tube
although there is no intrinsic string formation in this case,
as displayed in Fig.~\ref{f1.1}.
In the following we will present our
studies about the $q\bar q$ flux distributions
 which would support the above description.

  The remaining parts of this paper are arranged
as follows. Section~\ref{sec: Mea} gives the basic
concepts of the flux measurements.
Section~\ref{sec: Ftem} discusses the $q\bar q$ flux
distributions at finite temperatures and finite volume effects.
Section~\ref{sec: Fvol} discusses the relation between
the $q\bar q$ flux
distributions in a finite box and the string tension
in intermediate and large volumes. Finally,
Section~\ref{sec: Sum} gives the summary.

\bigskip
\section{ Measurements of The Flux Distributions }
\label{sec: Mea}

  In our study of the $SU(2)$ lattice gauge theory
(LGT) we use the standard Wilson action,
\begin{equation}
 S(U)=\beta \sum_P (1- \frac {1}{2} Tr U_P ),
\label{e2.1}
\end{equation}
where $U_P$ is the product of link variables around a
plaquette, i.e. $U_P(n)=U_\mu (n) U_\nu (n+\mu) U_{\mu}^{-1}(n+\nu)
U_{\nu}^{-1}(n)$.
We can measure the flux distributions
of a $q\bar q$ pair by calculating the quantity \cite{wh},
\FL
\begin{equation}
 f_{\mu \nu}(\vec r, \vec x)=\frac{\beta}{a^4}
\biggl [\frac{<P(0)P^{\dagger}(\vec r)\Box_{\mu \nu}(\vec x)>}
{<P(0)P^{\dagger}(\vec r)>}
-<\Box_{\mu \nu}>\biggr ],
\label{e2.2}
\end{equation}
   where $P(\vec r)\equiv \frac {1}{2}
 Tr \prod _{\tau=1}^{N_a} U_a ({\vec r}, \tau)$,
is the Polyakov loop closed in the $N_a$ direction and
$\Box_{\mu \nu}=\frac {1}{2} Tr(U_P)$ is the
plaquette variable with the
orientation $(\mu,\nu)$, which has 6 different values,
 $(\mu,\nu)$=$(2,3)$, $(1,3)$, $(1,2)$, $(1,4)$, $(2,4)$, $(3,4)$.

To reduce the fluctuations of the quantity
$P(0)P^{\dagger}(\vec r)\Box$, in practical calculations we
measure the quantity \cite{wh}
\FL
\begin{equation}
 f'_{\mu \nu}(\vec r, \vec x)=\frac{\beta}{a^4}
\biggl [\frac{<P(0)P^{\dagger}(\vec r)\Box_{\mu \nu}(\vec x)> -
<P(0)P^{\dagger}(\vec r)\Box_{\mu \nu}({\vec x}_R)>}
{<P(0)P^{\dagger}(\vec r)>} \biggr ],
\label{e2.2a}
\end{equation}
as the flux distribution instead of Eq.~(\ref{e2.2}),
where the reference point ${\vec x}_R$ was chosen to be
far from the $q\bar q$ sources. This
replacement does not change the measured average due to the
cluster decomposition theorem.
 The six components of $f'_{\mu \nu}$ in Eq.~(\ref{e2.2a}) correspond to
the components of the chromo-electric and chromo-magnetic fields
$(\vec {\cal E}, \vec {\cal B})$ in Minkowski space, i.e.
\begin{equation}
 f'_{\mu \nu}\rightarrow \frac {1}{2}
(-{\cal B}_1^2, -{\cal B}_2^2, -{\cal B}_3^2, {\cal E}_1^2,
{\cal E}_2^2, {\cal E}_3^2).
\label{e2.3}
\end{equation}

We then define the total electric energy density to be
\begin{equation}
 \rho_{el}=\frac {1}{2} [{\cal E}_1^2 +
   {\cal E}_2^2 + {\cal E}_3^2 ],
\label{e2.4}
\end{equation}
and the total magnetic energy density
\begin{equation}
 \rho_{ma}=\frac {1}{2} [{\cal B}_1^2 +
   {\cal B}_2^2 + {\cal B}_3^2 ].
\label{e2.5}
\end{equation}
   In the following we will concentrate on studying the
total energy and action densities $\rho_E$ and $\rho_A$,
which are the combinations of $\rho_{el}$ and $\rho_{ma}$,
The total energy density is
\begin{equation}
 \rho_E=\rho_{el}+ \rho_{ma},
\label{e2.6}
\end{equation}
The total action density is
\begin{equation}
\rho_A=\rho_{el}- \rho_{ma}.
\label{e2.7}
\end{equation}

In our measurements we transformed our
flux data from lattice units to physical units by using the
scaling relation given by Table ~\ref{t2.1}, which were obtained
from a similar table in ref.\cite{hw} and we interpolated
points in the region $2.22< \beta <2.50$. The detail of
the interpolation process is presented elsewhere\cite{ph}.

\bigskip
\section{ Flux Distributions at Finite Temperatures }
\label{sec: Ftem}

  It is well known that $SU(N)$ gauge theory has a deconfining
phase transition at some finite temperature, $T_c$ \cite{ps}. This
finite temperature phase transition has been studied extentively
in $SU(N)$ LGT. The transition temperature $T_c$ can be
determined from Monte Carlo calculations, for example,
the critical temperature for $SU(2)$ was determined to be \cite{efw}
\begin{equation}
 \beta_c=2.2985 \pm 0.0006 \qquad \mbox{(for $N_a=4$)}.
\label{e3.1}
\end{equation}
The relation between $\beta_c$ and $T_c$ is,
$T_c=1/N_a a(\beta_c)$, as given by Eq.~(\ref{e1.4}).
The lattice spacing $a$ is a function of the coupling
constant $\beta$, such as that of Table ~\ref{t2.1}. The phase
transition is expected to occur in the infinite volume limit,
i.e. $N_b$, $N_c \rightarrow \infty $. In this limit, for
$\beta < \beta_c$, the system is in the confined phase,
otherwise, it is in the unconfined phase.

  We measured the $q\bar q$ flux distributions on lattices
$N_a \cdot N_b^2 \cdot N_c$ with $N_a=4$, $N_b=5$, 7, 9 and 11,
and $N_c=65$. Since $N_c\gg N_a$, $N_b$, as $N_b/N_a$ gets large,
we expect the system to approach the infinite volume limit and
one should see the two phases.
However, for $N_b$ small, that is, $N_b$ satisfying the condition,
\begin{equation}
 N_b\simeq N_a, \qquad \mbox {and} \quad N_b a/r \le 1,
\label{e3.0}
\end{equation}
 the $q\bar q$ pair is in a finite box and one expects finite
volume effects to be large. In the following we will study
the flux distributions with $\beta > \beta_c$ and
$\beta < \beta_c$ in various spacial volumes respectively.

\bigskip
\subsection{ Flux Distributions With $\beta > \beta_c$ }
\label{subsec: blbc}

  In this case the $q\bar q$ system approaches
the unconfined phase as the volume becomes large
($N_b \rightarrow \infty $). In this phase we expect that
 there is no string formation. However, for small volume
one expects finite volume effects to be large.

In Fig.~\ref{f3.1} we show some typical results of the energy
density $\rho_E$ distributions in the region of $\beta > \beta_c$
($\beta=2.40$). They are the flux distributions on the transverse
plane midway between the $q\bar q$ sources with fixed separation
$r=4a$, and they were measured on lattices of 4 different spatial size
$N_b$=5, 7, 9, 11. For $N_b$ small (i.e. $N_b$=5, 7) the transverse
plane is the whole lattice. For large $N_b$ ($N_b$=9, 11)
the data is truncated on the margins. The signal is lost in the
noise beyond the region shown.

   From Fig.~\ref{f3.1} we can compare the flux distributions
in boxes of different transverse size $N_b$. One can clearly
see for $N_b$ small the flux density $\rho_E$ at the edges of
each plane have large values, as shown in Fig.~\ref{f3.1} (a), (b).
This implies that finite volume effects are significant when $N_b$
is small because in small volumes flux lines of the $q\bar q$ sources
would be squeezed dramatically by side walls of the box, so the
values of $\rho_E$ at edges are large in this case.
 As we increase the transverse size, $N_b$,
finite volume effects become smaller, the flux density $\rho_E$
at edges decrease rapidly to zero, as shown in
Fig. ~\ref{f3.1} (c), (d).

   Here we want to emphasize that the values of $\rho_E$
at the edges of each plane are not due to the reference point
in Eq.~(\ref{e2.2a}).
In our flux measurement we choose the reference point ${\vec x}_R$ far
from the $q\bar q$ sources, and we find the reference value
$<P(0)P^{\dagger}(\vec r)\Box_{\mu \nu}({\vec x}_R)>$ in
Eq.~(\ref{e2.2a}) is consistent with the product
$<P(0)P^{\dagger}(\vec r)><\Box_{\mu \nu}>$ within errors.

  In Table ~\ref{t3.1} we list the typical values of flux
densities $\rho_E$, $\rho_A$ at edges of each plane in
Fig.~\ref{f3.1} and their corresponding errors.
{}From this table one can explicitely see that the values $\rho_E$
and $\rho_A$ at edges decrease rapidly with $N_b$,
as we observed from Fig.~\ref{f3.1}.
 However, we also notice that even in cases of large $N_b$
(i.e. $N_b$=9, 11) the values of $\rho_A$ shown are
non-zero within errors.
This may be caused by a number of factors. The edge of the
plane is not the boundary of the lattice for cases of
$N_b$=9, 11, the flux values
may in fact vanish on the boundary of the lattice,
but we did not calculate them there for practical reasons.
Also $N_b$ is perhaps not
large enough (even for $N_b=11$), so there are
still some small finite volume effects. Finally our
error bars were only calculated from statistical error, the
actual errors may be larger due to systematic errors.

   To see the behaviour of the flux distributions changing with the
$q\bar q$ separation $r$, in Fig.~\ref{f3.3} we show the
$\rho_A$ distribution on the transverse plane for 4 different
separations, $r=3a$, $4a$, $5a$ and $6a$. The flux data were
measured on the lattice $4\cdot 5^2\cdot 65$ with $\beta =2.40$.
{}From this figure one can see the peak values of $\rho_A$
decrease rapidly with the increase of $r$. At large $r$
(i.e. $r=6a$) the peak of $\rho_A$ almost vanishes,
the flux density on the plane approachs a uniform distribution.

 We then calculated the center
slice energy $\sigma_E$ and action $\sigma_A$ from our
flux data, which are the energy and action stored in the
transverse slice of unit thickness midway between the $q\bar q$ pair.
The results are shown in Fig.~\ref{f3.5}.
 In this figure we plot the behaviours of $\sigma_E$ vs. $r$
and $\sigma_A$ vs. $r$ respectively for $\beta=2.36$. This
shows that for $N_b$ small $\sigma_E$ and $\sigma_A$
do not decrease to zero as $r$ increaces.
However, for $N_b$ large (e.g. $N_b=9$, 11), $\sigma_E$
and $\sigma_A$ decrease rapidly with $r$ and become
very small at large $q\bar q$ separations (i.e. $r=6a$).

\bigskip
\subsection{ Analysis Of Finite Volume Effects For Cases
With $\beta > \beta_c$ }
\label{subsec: afeblbc}

    To see how the finite volume effects influence
 $q\bar q$ flux distributions in a finite box,
let us consider an electrostatic charge pair
 $+e$, $-e$ enclosed
in a similar long rectangular box as that of Fig.~\ref{f1.1}.
The interaction between charges is the Coulomb interaction
$V(r)\sim 1/r$. As the charge separation $r$ becomes very
large, i.e. $r/N_b a \rightarrow \infty$,
one can assume the electric field $\vec {\cal E}$ on the middle
transverse plane is uniform, and can be written as
\begin{equation}
 {\cal E}=\frac {\Phi}{(N_b a)^2},
\label{e3.2}
\end{equation}
where $\Phi$ is electric flux through the transverse plane,
which is a constant for the Coulomb interaction, and $(N_b a)$ is
the transverse size of the box. So the total electric field
energy on the transverse plane is
\begin{equation}
 (\sigma_E)_C=\frac {1}{2}{\cal E}^2 (N_b a)^2
  \sim \frac {1}{(N_b a)^2}.
\label{e3.3}
\end{equation}
This shows that at large charge separations
the center slice energy $(\sigma_E)_C$ decrease
with the behaviour $(N_b a)^{-2}$ as the transverse size $N_b$
increases, and $(\sigma_E)_C$ vanishes as $N_b\rightarrow \infty$,
where the label `C' denotes Coulomb interaction.

Now let us return to the $q\bar q$ pair in the box with
$\beta > \beta_c$. As we discussed in the above section, the
system approaches the unconfined phase for large $N_b$.
In this phase the $q\bar q$ interaction is a screened
Coulomb interaction $V(r)\sim e^{-mr}/r$ with $m$ the
screening mass \cite{kps}. Then if the $q\bar q$ pair is put in a box
such as that in Fig. ~\ref{f1.1}, the flux $\Phi$ through the transverse
plane is not a constant, but would decrease with $r$
and $N_b$. So we expect that in this case the
center slice energy $\sigma_E$ for a $q\bar q$ pair
would decay faster with $N_b$ than $(\sigma_E)_C$,
which has the inverse square behaviour,
$(\sigma_E)_C \sim (N_b a)^{-2}$, for $r\rightarrow \infty$.
In Fig.~\ref{f3.7} we plot our $\sigma_E$ data versus the
transverse size $N_b$ for large $q\bar q$ separation $r$.
Our data are compared with the Coulomb behaviour $(N_b a)^{-2}$
and for reference purpose the inverse quartic behaviour
$(N_b a)^{-4}$. The data
were measured on lattices $4\cdot N_b^2\cdot 65$ with
$N_b=$5, 7, 9, 11 and $\beta=2.40$.

  From Fig.~\ref{f3.7} one can see that for large $r$ our $\sigma_E$
data appears to decay faster with $N_b$ than the Coulomb behaviour
$(N_b a)^{-2}$ as expected. This shows us that the
$q\bar q$ interaction in the unconfined phase
at least contains a term that decays faster than the
Coulomb interaction, such as the screened Coulomb interaction,
although the data is not good enough to determine the
screening mass.

  In conclusion, our flux data in the region of $\beta > \beta_c$
shows that in the unconfined phase there is no string formation.
For small transverse size $N_b$, the finite volume effects
are large, the flux lines between a $q\bar q$ would be squeezed by
side walls of the box significantly. This would
result in a finite string tension.

\bigskip
\subsection{ Flux Distributions With $\beta < \beta_c$ }
\label{subsec: bsbc}

 For $\beta < \beta_c$ the $q\bar q$ system approaches
the confined phase as the volume becomes large
($N_b\rightarrow \infty$). We expect
string formation would occur in this phase. To see this
we need to study behaviours of the flux distributions
as a function of the $q\bar q$ separation $r$.

  In Fig.~\ref{f3.11} we show the action density
$\rho_A$ distribution changing with the $q\bar q$ separation $r$.
The flux data were measured on the lattice
$4\cdot 11^2\cdot 65$ with $\beta =2.25$.
Since $N_b$ is large ($N_b$=11) we deleted the
margins of the transverse cross
section of the lattice, as we did in Fig.~\ref{f3.1}.
This figure should be compared with Fig.~\ref{f3.3}, which
is for $\beta > \beta_c$. One can see Fig.~\ref{f3.11}
shows significantly different behaviour from Fig.~\ref{f3.3}.
In Fig.~\ref{f3.11} the peak values of $\rho_A$ on the plane
approach a finite value as $r$ becomes large.
Even at large $r$ (i.e. $r=6a$) this peak still exists. However,
in Fig.~\ref{f3.3} the peak of the $\rho_A$ distribution almost
disappears at $r=6a$. So the flux distribution in Fig.~\ref{f3.11}
implies that intrinsic string formation occurs in the region
of $\beta < \beta_c$.
This is not due to finite volume effects because these effects
are small at large volumes (i.e. $N_b$=9, 11), as we discussed
in Fig.~\ref{f3.1} and Table~\ref{t3.1}.

 We also calculated the center slice energy and action
$\sigma_E$, $\sigma_A$ from our flux data in the region
$\beta < \beta_c$. If string formation occurs in the confined
phase, for $N_b$ large both $\sigma_E$ and $\sigma_A$
should approach to some finite non-zero constants when
$r\rightarrow \infty$.
In Figs.~\ref{f3.12} we plot the behaviours of
$\sigma_E$ vs. $r$ and $\sigma_A$ vs. $r$ respectively
for $\beta=2.28$. The data were measured on lattices of
various spatial size $N_b$=5, 7, 9, 11.
{}From this figure one can see that in all cases $\sigma_E$ and
$\sigma_A$ do not decrease with $r$. For each fixed $N_b$
the values of $\sigma_E$ and $\sigma_A$ are almost
constant as $r$ increases. Further, for large $N_b$
($N_b=9$ ,11), where finite volume effects are small,
both $\sigma_E$ and $\sigma_A$ keep as
finite non-zero constants as $r$ becomes large.
This behaviour is totally different from that of Fig.~\ref{f3.5},
which is in the region of $\beta > \beta_c$.
So Fig.~\ref{f3.12} also implies that string formation occurs.
In this figure we also notice that fluctuations of the
data are large compared to the unconfined data. This is a
typical behaviour because confinement corresponds to disorder.

In conclusion, our flux data in the region of $\beta < \beta_c$
provide evidence for intrinsic string formation in the confined phase.
This string formation is not due to finite volume effects
because these effects are small as the volume becomes
large (i.e. $N_b=9, 11$).

\bigskip
\section{$SU(2)$ Gauge Theory In Finite Volumes At $T_B\approx 0$}
\label{sec: Fvol}

   As we discussed in section~\ref{sec: intro}, we have two
ways to interpret the LGT calculations. One way is to identify
the shortest extent $N_a$ of lattices to be the temporal size.
This is a convenient way to study the LGT system. In previous
sections we have discussed our calculation results in this
way. However, to compare with analytical
results in finite volumes at zero temperatures, one can use
another way to interpret LGT results, that is, the longest
extent $N_c$ of lattices is chosen as the time direction,
so that the temperature is as low as possible.
In this case Polyakov loops closed in $N_a$ (or $N_b$)
direction are no longer viewed as quark sources,
they are considered to be spatial operators.
Quantities such as, glueball mass and string tension,
can be calculated from Polyakov loop correlations along
the time direction ($N_c$). No matter which way we choose,
quantities calculated in LGT are the same, the mathematics
of the two ways are equivalent. In this section we shall
look at the system in terms of this second interpretation,
and reinterpret the results of previous sections.

  If we choose $N_c$ as the time direction,
we are studying $SU(2)$ gauge theory in the volume
$U=N_a\cdot N_b^2\cdot a^3$ at near zero temperature
$T_B=1/N_c a$ because $N_c$ is large. For convenience,
instead of $z_g$ defined in Eq.~(\ref{e1.1})
we shall use another parameter,
$z_{\kappa}={\sqrt \kappa}L={\sqrt \kappa}N_b a$, to
characterize the physical size of the volume \cite{bb}, where
$\kappa$ is the string tension measured by two
Polyakov loops closed in $N_a$ with correlations measured
along $N_c$. This corresponds to $\kappa_t$ of ref.\cite{bb}.
J. Koller and P. van Baal used $z_g$ in their analytical
calculations \cite{vk}, so that they could easily compare with
Monte Carlo results. B.A. Berg and A.H. Billoire used both
$z_g$ and $z_{\kappa}$ parameters and other $z$ parameters
in analyses of their LGT Monte Carlo data for convenience \cite{bb}.
The parameter $z_\kappa$ is equivalent to $z_g$,
For example, from the data of ref.\cite{bb}
one can see, $z_g\approx 1$ corresponds to $z_{\kappa}\approx 0.24$,
and $z_g\approx 5$ corresponds to $z_{\kappa}\approx 1.3$.
In Table~\ref{t4.1} we show the correspondence of $z_g$ and
$z_\kappa$.

\bigskip
\subsection{Results In Intermediate Volumes ($0.24 < z_\kappa < 1.3$) }
\label{subsec: rintv}

  In this part we will show that our lattices with $\beta > \beta_c$
belong to the intermediate volume region.
 Since in this region our string tension data have large
error bars, we just simply use the string tension data of
ref.\cite{bb} to show that our lattices with
$\beta > \beta_c$ satisfy $0.24< z_\kappa < 1.3$.
The data of ref.\cite{bb} are given in Table~\ref{t4.2},
which have high statistical accuracies.

    From Table~\ref{t4.2} we can see the values of string tension
${\sqrt \kappa}a$
and the parameter $z_\kappa$ decrease with the increase
of $N_b$ and $\beta$. Our data were measured on lattices of
the size $4\cdot N_b^2 \cdot 65$ with $N_b=$5, 7, 9, 11 and
$\beta=2.36$ and 2.40, which are similar to the lattices
 in Table~\ref{t4.2}.
We can estimate our string tension and values of
$z_\kappa$ from Table~\ref{t4.2}. For example,
the estimated string tension for the lattice
$4\cdot 5^2 \cdot 65$ with $\beta=2.36$ could be interpolated from
the string tension measured on lattices $4\cdot N_b^2 \cdot 64$
with $N_b$=4, 6 and $\beta=2.36$. From Table~\ref{t4.2} one can
see that for $N_b$=5 the string tension satisfies,
$0.1593 < {\sqrt \kappa}a < 0.2475$, and
$0.80 < z_\kappa={\sqrt \kappa}a N_b < 1.24$,
which is in the intermediate volume region, $0.24 < z_\kappa < 1.3$,
as shown in Table~\ref{t4.1}.
By comparing our lattices with Table~\ref{t4.2} in this way,
we find that our lattices with $\beta=2.36$ and 2.40 for
$N_b=$5, 7, 9, 11 are all in the intermediate volume region,
with the ``{\it box temperature}" approaching zero,
i.e. $T_B\rightarrow 0$.

  As we discussed in section~\ref{sec: Ftem}, for
$\beta > \beta_c$ there is no intrinsic string formation.
However, since finite volume effects are large when the
transverse size $N_b$ is small, this results in the observed
string tension in these cases. As $N_b$ becomes larger,
finite volume effects becomes smaller, the string tension
becomes smaller. This is confirmed in Table~\ref{t4.2}.
Since the lattices with $\beta > \beta_c$ in our study
all belong to the intermediate volume region, in these
cases we find there is no intrinsic string formation,
and the string tension is due to finite volume effects.
In general, we expect that the results apply to the
whole intermediate volume region in the zero temperature
limit ($T_B\rightarrow 0$).

\bigskip
\subsection{Results In Large Volumes ($z_\kappa > 1.3$) }
\label{subsec: rlarv}

 For $\beta < \beta_c$ we can easily extract the string tension,
which is given in Table~\ref{t4.3}. In this table we also show
the values of $z_\kappa$ for lattices with $\beta < \beta_c$.
   We can see that most lattices listed in Table~\ref{t4.3}
satisfy $z_\kappa > 1.3$, with one case in the critical region,
$z_\kappa \approx 1.3$ for $\beta=2.28$ and $N_b=5$. From
Table~\ref{t4.1} we know that lattices satisfying
the condition, $z_\kappa > 1.3$,
belong to the large volume region in the zero temperature
limit ($T_B\rightarrow 0$). So our lattices with
$\beta < \beta_c$, as listed in Table~\ref{t4.3}, are in this
region.

 As we discussed in section~\ref{subsec: bsbc},
a system with $\beta < \beta_c$ is in the confined phase
in the infinite volume limit. In this case we have shown that
intrinsic string formation occurs.
As the transverse size $N_b$ is made small (e.g. $N_b=5$),
we also observed some finite volume effects,
however, it does not change the nature of $q\bar q$ confinement.
We have shown that the lattices with
$\beta < \beta_c$ in our study all belong to the
large volume region in the zero temperature
limit ($T_B\rightarrow 0$). From our study we have found
evidence for intrinsic string formation in these cases,
which is not due to finite volume effects. We expect that
the intrinsic string formation occurs in the whole
large volume region.

\bigskip
\section{Summary}
\label{sec: Sum}

   We used two ways to interpret the LGT results calculated
on lattices $N_a \cdot N_b^2 \cdot N_c$ with the geometry
$N_a \leq  N_b \ll N_c$. First, we chose the convenient way
to interpret our data, by identifying $N_a$ to be the
time direction we can study the system at finite temperatures
($T_p=1/N_a a$) in the volume $V=N_b^2\cdot N_c\cdot a^3$.
We analysed the flux distributions of a $q\bar q$ pair in
terms of this interpretation.
If we choose $N_c$ to be the time direction we can
study the system at near zero temperature ($T_B=1/N_c a$)
in the volume $U=N_a\cdot N_b^2\cdot a^3$.
We find that for $\beta > \beta_c$
there is no intrinsic string formation. Our lattices with
$\beta > \beta_c$ belong to the intermediate volume region.
We then established that the origin of the string tension
measured by Berg and Billoire \cite{bb} in this region
is related to finite sizes of the lattice.
For $\beta < \beta_c$ we find clear signals for
intrinsic string formation. Our lattices with
$\beta < \beta_c$ are in the large volume region,
but near the borderline of intermediate volumes and
large volumes. This is just beyond the volume region
investigated by Berg and Billoire.

  Another remark on string formation is that for higher units
of 't Hooft electric flux one expects that string formation predicts
the relation \cite{bb},
\begin{equation}
 E_n =\sqrt {n} E_1 \qquad (n=1,2, \cdots ),
\label{e5.1}
\end {equation}
  where $E_n$ is the energy of the $n$ unit flux. However,
both analytical results\cite{vk} and LGT calculations\cite{bb}
produce a different behaviour in the intermediate volume region,
\begin{equation}
 E_n \approx n E_1,
\label{e5.2}
\end {equation}
 where $n=1, 2, 3$ for $SU(2)$. Even in the large volume
region some Monte Carlo data of ref.\cite{bb} still show
the behaviour of Eq.~(\ref{e5.2}). We speculate that
Eq.~(\ref{e5.1}) may be due to lattice artifacts since
the origin of $\sqrt {2}$ and $\sqrt {3}$ in this equation
are for the planar diagonal and volume diagonal on lattices.

Our study supplement the global study of Berg and Billoire\cite{bb},
and provide evidence that string formation does not occur
in intermediate volumes, but occurs in large volumes.
To verify Eq.~(\ref{e5.1}) or Eq.~(\ref{e5.2}) we need
to measure the energy $E_n$ of higher unit flux in the
large volume region.

\bigskip
\section*{Acknowledgments}

  We wish to thank P. van Baal, B.A. Berg, A.H. Billoire, J. Wosiek,
D.A. Browne and V. Singh for many fruitful discussions on
this problem. This research is supported by the U.S. Department of
Energy under grant DE-FG05-91ER40617.

\newpage

\figure{ The $q\bar q$ colour sources in a long rectangular
box. The flux lines between the $q\bar q$ pair are confined by
side walls and have a flux tube form. \label{f1.1} }

\figure{ The energy density $\rho_E$ flux distributions
in the region of $\beta > \beta_c$ on the plane midway between the
$q\bar q$ sources of separation $r=4a$. The flux
data were measured on lattices
$4\cdot N_b^2 \cdot 65$ with $\beta=2.40$ for various
spatial sizes, (a) $N_b=5$, (b) $N_b=7$, (c) $N_b=9$,
and (d) $N_b=11$. These flux data are measured in the physical unit
$Gev/fm^3$. \label{f3.1}}

\figure{ The action density $\rho_A$ flux distributions
in the region $\beta > \beta_c$ on the plane midway between the
$q\bar q$ sources of various separations (a) $r=3a$,
(b) $r=4a$, (c) $r=5a$ and (d) $r=6a$.
The flux data were measured on lattices
$4\cdot 5^2 \cdot 65$ with $\beta=2.40$,
The data are measured in the physical unit
$Gev/fm^3$. \label{f3.3}}

\figure{ The behaviour of center slice energy
$\sigma_E$ and action $\sigma_A$ verses the $q\bar q$
 separaction $r$ in the region of $\beta > \beta_c$,
(a) $\sigma_E$ vs. $r$, (b) $\sigma_A$ vs. $r$. The data
were measured on lattices $4\cdot N_b^2\cdot 65$ with
$\beta=2.36$ for various spatial size,
 $N_b=5$ (circles), $N_b=7$ (squares), $N_b=9$ (triangles),
$N_b=11$ (diamonds). The data are
in the physical unit $Gev/fm$. \label{f3.5}}

\figure{ The plot of $\sigma_E$ in the region of $\beta > \beta_c$
verses the transverse
size of lattices $N_b$ at large $q\bar q$ separation
$r=6a$. The solid line is the Coulomb behaviour $(N_b a)^{-2}$,
the dashed line is the inverse quartic behaviour $(N_b a)^{-4}$,
both are normalized to the data at $N_b=5$. The data were measured
on lattices $4\cdot N_b^2\cdot 65$ with $N_b$=5, 7, 9, 11.
and $\beta=2.40$. \label{f3.7}}

\figure{ The action density $\rho_A$ flux distributions
in the region of $\beta < \beta_c$ on the plane midway between the
$q\bar q$ sources of various separations (a) $r=3a$,
(b) $r=4a$, (c) $r=5a$ and (d) $r=6a$.
The flux data were measured on lattices
$4\cdot 11^2 \cdot 65$ with $\beta=2.25$,
The data are measured in the physical unit
$Gev/fm^3$. \label{f3.11}}

\figure{ The behaviour of center slice energy
$\sigma_E$ and action $\sigma_A$ verses the $q\bar q$
 separaction $r$ in the region of $\beta < \beta_c$,
(a) $\sigma_E$ vs. $r$, (b) $\sigma_A$ vs. $r$. The data
were measured on lattices $4\cdot N_b^2\cdot 65$ with
$\beta=2.28$ for various spatial size,
 $N_b=5$ (circles), $N_b=7$ (squares), $N_b=9$ (triangles),
$N_b=11$ (diamonds). The data are
in the physical unit $Gev/fm$. \label{f3.12}}

\mediumtext

\begin{table}
\caption{ The correspondence of the lattice spacing
$a$ and the coupling constant $\beta$ for $SU(2)$ LGT.}
\begin{tabular}{ccccc}
$\beta$ & \dec 2.25  & \dec 2.28  & \dec 2.36  & \dec 2.40  \\
$a$ ($fm$)  & \dec 0.1966  & \dec 0.1785  & \dec 0.1374  & \dec 0.1203 \\
\end{tabular}
\label{t2.1}
\end{table}
\bigskip

\begin{table}
\caption{ The typical measured flux density values (errors) of
$\rho_E$ and $\rho_A$ in the region of $\beta > \beta_c$
on the edges of the transverse plane midway between the
$q\bar q$, as shown in Fig.~\ref{f3.1}.
The flux data are in the physical
unit $Gev/fm^3$.  }
\begin{tabular}{cccccc}
\multicolumn{6} {c} {$4\cdot N_b^2 \cdot 65$}  \\
\tableline
 $r=4a$  & $N_b$  &   5   &   7   &   9   &   11    \\
\tableline
$\beta=2.40$ & $\rho_E$ & 2.5 (5)  & 0.86 (24)  &  0.11 (18)
                        & 0.31 (17)    \\
             & $\rho_A$ & 17.0 (6) & 2.9 (3)  &  1.5 (2)
                        & 1.0 (2)    \\
\end{tabular}
\label{t3.1}
\end{table}
\bigskip

\begin{table}
\caption{The correspondence of the two parameters
 $z_g= m(0^+)N_b a$ and $z_{\kappa}={\sqrt \kappa}N_b a$,
 which are obtained from data in ref.\protect\cite{bb}.}
\begin{tabular}{ccc}
small volume &  intermediate volume &  large volume \\
\tableline
$z_g<1$    &  $1<z_g<5$         &  $z_g>5$    \\
$z_\kappa < 0.24$ & $0.24 < z_\kappa < 1.3$ & $z_\kappa > 1.3$ \\
\end{tabular}
\label{t4.1}
\end{table}
\bigskip

\begin{table}
\caption{The values of the string tension data and $z_\kappa$
on lattices of size $4\cdot N_b^2 \cdot 64$ with
$N_b=$4, 6, 8 and $\beta=2.36$, 2.38 and 2.41. The
data are quoted from ref.\protect\cite{bb}. }
\begin{tabular}{ccccc}
\multicolumn{5} {c} {$4\cdot N_b^2 \cdot 64$}  \\
\tableline
\multicolumn{2} {c} {$N_b$}  & $4$  &  $6$ & $8$  \\
\tableline
$\beta=2.36 $ &  $ {\sqrt \kappa}a$ &  0.2475 (18)
              & 0.1593 (39) & 0.1091 (39)   \\
              &  $z_\kappa =N_b {\sqrt \kappa}a$ &  0.99 (1)
              & 0.96 (2) &  0.87 (3)     \\
\tableline
$\beta=2.38 $ &  $ {\sqrt \kappa}a$ &  0.2424 (23)
              & 0.1357 (44)  & 0.0732 (81)   \\
              &  $z_\kappa =N_b {\sqrt \kappa}a$ & 0.97 (1)
              & 0.81 (4) &  0.59 (6)   \\
\tableline
$\beta=2.41 $ &  $ {\sqrt \kappa}a$ &  0.2264 (14)
              & 0.1314 (51)  &        \\
              &  $z_\kappa =N_b {\sqrt \kappa}a$ & 0.91 (1)
              & 0.79 (3) &      \\
\end{tabular}
\label{t4.2}
\end{table}
\bigskip

\begin{table}
\caption{ The string tension data $\kappa$ and
 the values of the $z_\kappa$ parameter on lattices of the size
$4\cdot N_b^2 \cdot 65$ with $N_b=$5, 7, 9, 11 and
$\beta=2.25$ and 2.28.  }
\begin{tabular}{cccc}
\multicolumn{4} {c} {$4\cdot N_b^2 \cdot 65$}  \\
\tableline
 $N_b$ & $\beta$ & ${\sqrt \kappa}a$
        & $z_\kappa =N_b {\sqrt \kappa}a$     \\
\tableline
  5    & 2.25   &  0.300 (22)  &  1.5 (1)     \\
       & 2.28   &  0.251 (34)  &  1.3 (2)     \\
\tableline
  7    & 2.25   &  0.288 (31)  &  2.2 (2)     \\
       & 2.28   &  0.258 (31)  &  1.8 (2)     \\
\tableline
  9    & 2.25   &  0.262 (84)  &  2.4 (8)     \\
       & 2.28   &  0.226 (67)  &  2.0 (6)     \\
\tableline
  11   & 2.25   &  0.307 (48)  &  3.4 (5)     \\
       & 2.28   &  0.232 (64)  &  2.6 (7)     \\
\end{tabular}
\label{t4.3}
\end{table}

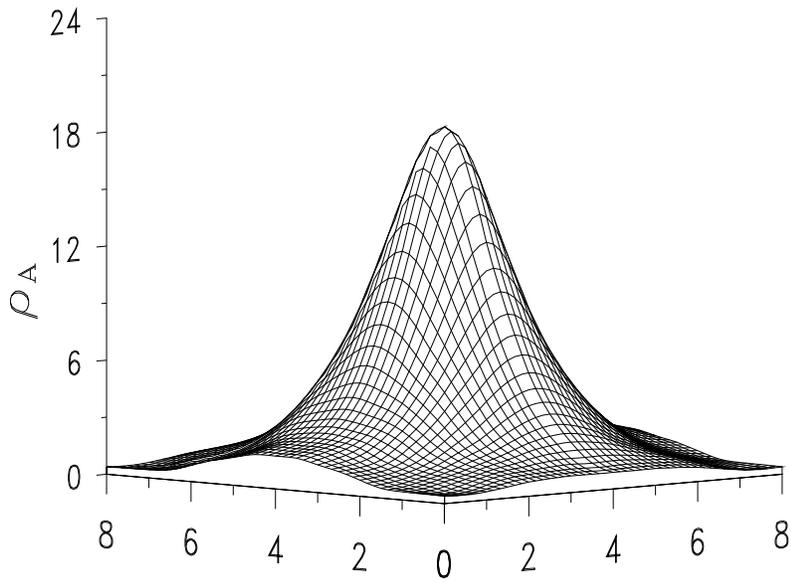
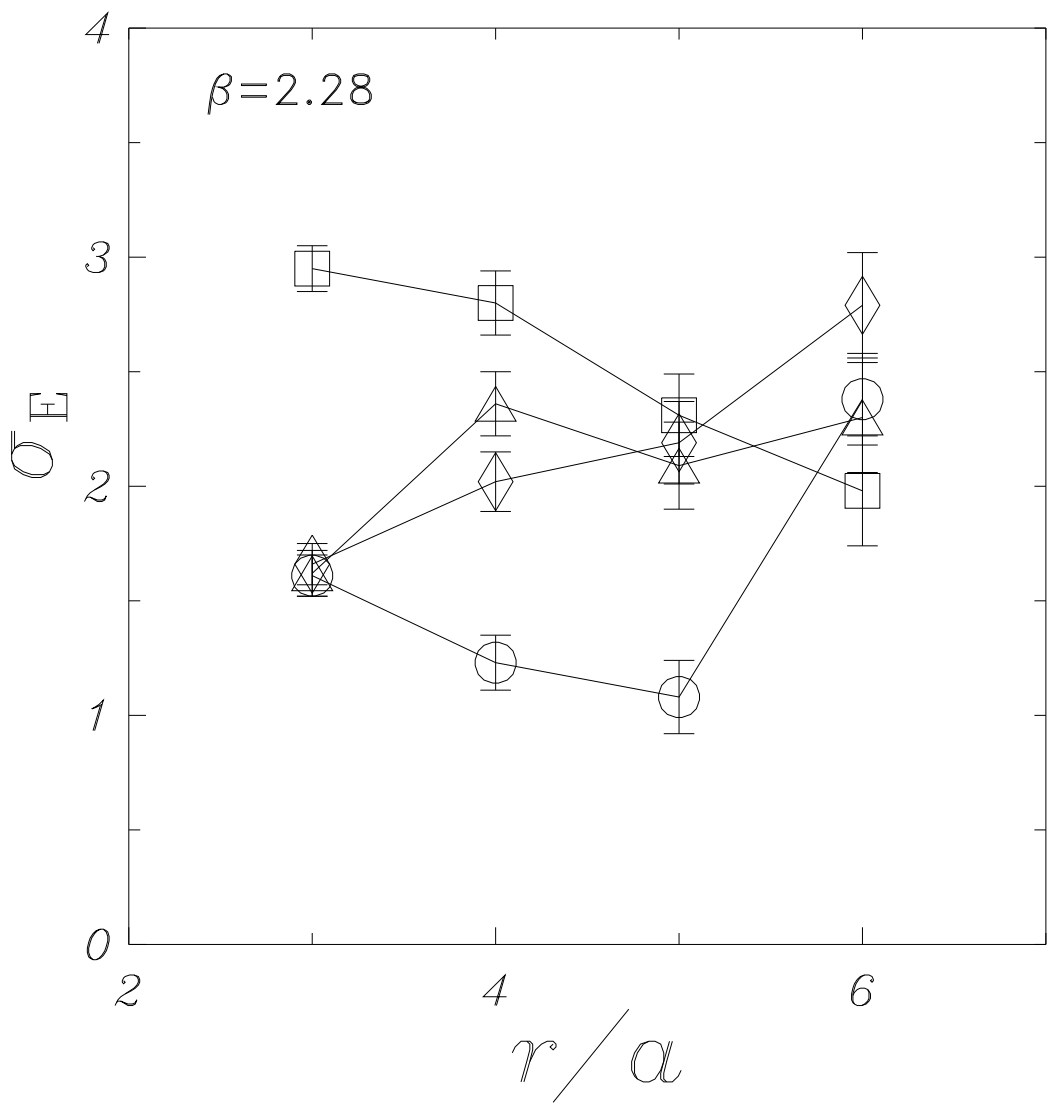
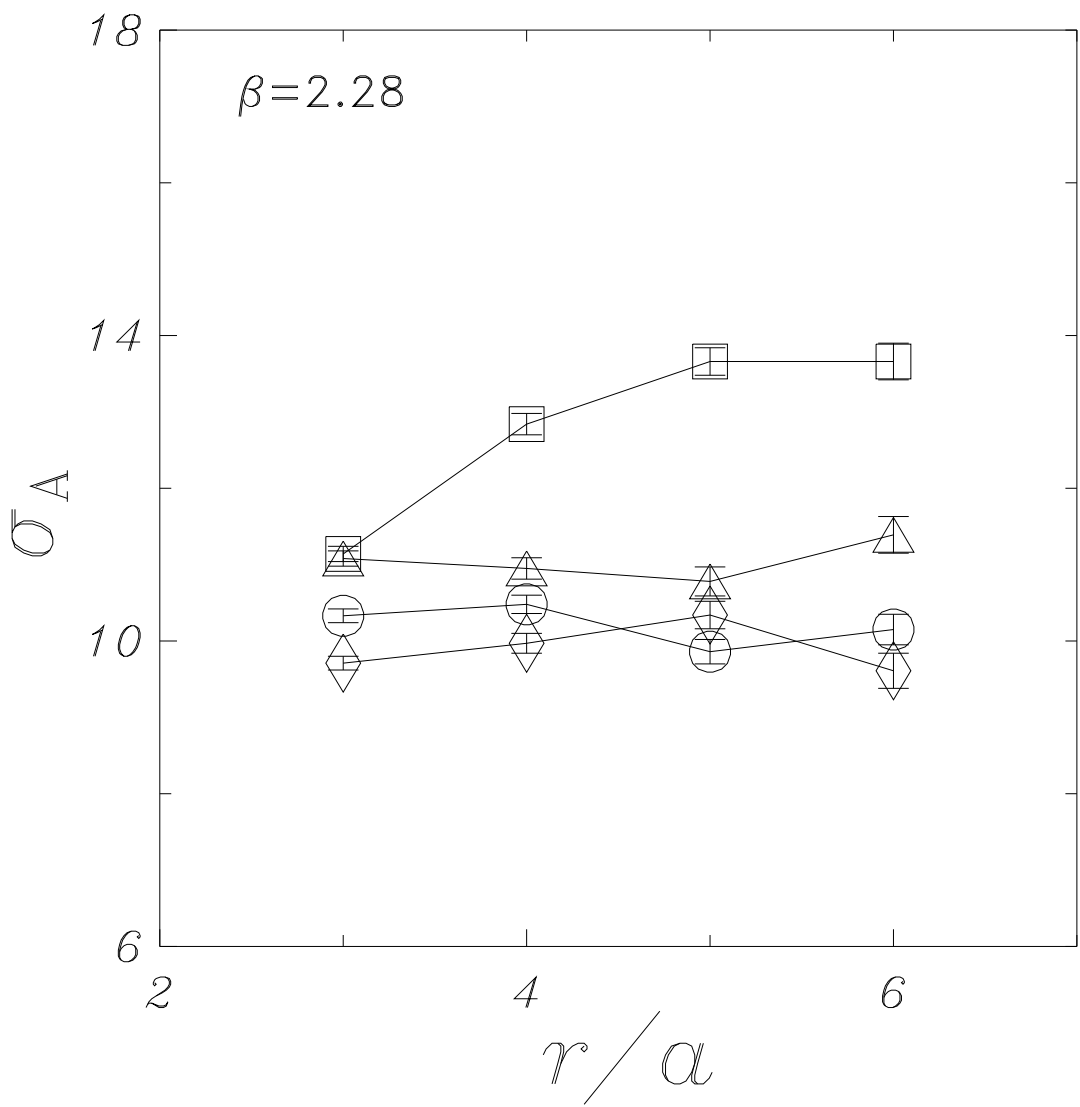
\end{document}